# Online Learning of Interconnected Neural Networks for Optimal Control of an HVAC System

Young-Jin Kim

*Abstract*—Optimizing the operation of heating, ventilation, and air-conditioning (HVAC) systems is a challenging task, requiring the modeling of complex nonlinear relationships among HVAC load, indoor temperatures, and outdoor environments. This paper proposes a new strategy for optimal operation of an HVAC system in a commercial building. The system for indoor temperature control is divided into three sub-systems, each of which is modeled using an artificial neural network (ANN). The ANNs are then interconnected and integrated into an optimization problem for temperature set-point scheduling. The problem is reformulated to determine the optimal set-points using a deterministic search algorithm. After the optimal scheduling is initiated, the ANNs undergo online learning repeatedly, mitigating the overfitting. Case studies are performed to analyze the performance of the proposed strategy, compared to the strategies with a pre-determined temperature set-point, an ideal physics-based building model, and conventional ANN-based building models. The case study results confirm that the proposed strategy is effective in terms of the HVAC energy cost, practical applicability, and training data requirement.

## Nomenclature

The main notations used in this paper are summarized here.

**A. Sets and Indices:**

| | |
|---|---|
| $d, t$ | indices for day and time |
| $m, n$ | indies for neural networks and linear power blocks |
| $max, min, ref, set$ | subscripts for maximum, minimum, reference, and set-point values |
| $e(\cdot, \cdot')$ | normalized root mean square error between $\cdot$ and $\cdot'$ |
| $L_1, L_2, L_3$ | neural networks to model building thermal dynamics |
| $S_1, S_2, S_3$ | original and reformulated optimization problems |

**B. Parameters:**

| | |
|---|---|
| $C^t$ | retail electricity price at time $t$ |
| $D_T$ | dead-band between $T_{set}^t$ and $T_i^t$ |
| $\mathbf{E}^t$ | building thermal environments at time $t$ |
| $F_{n,\tau}^t$ | linear gradient of $T_i^t$ at time $t$ resulting from input power segment $n$ of HVAC system at time $\tau$ |
| $L_P$ | maximum time delay of input data for neural networks |
| $N_d$ | number of days for online supervised learning |
| $N_{ET}, N_{EO}$ | numbers of epochs for network training and optimal scheduling |
| $N_{HLm}$ | number of hidden layers of the $m_{th}$ network |
| $N_{HNm}$ | number of hidden nodes in each layer of the $m_{th}$ network |
| $N_{ID}$ | number of initial training datasets |
| $N_S$ | number of linear power blocks |
| $N_T$ | number of scheduling time intervals in a day |
| $P_{max}, P_{min}$ | maximum and minimum power inputs of HVAC system |
| $P_O$ | offset of reference power input to HVAC system |
| $Q_i^t$ | internal thermal load of a building at time $t$ |
| $R_H, R_L$ | upward/downward ramp rate limits of the power input to HVAC system |
| $R_T, R_O$ | learning rates for network training and optimal scheduling |
| $T_a^t$ | adjacent room temperature at time $t$ |
| $T_e^t$ | evaporator-side air temperature at time $t$ |
| $T_{in}^t$ | indoor temperature at time $t$ under the no cooling condition from 1 to $t$; i.e., the HVAC system remains off. |
| $T_{i,max}^t, T_{i,min}^t$ | maximum and minimum limits of $T_i^t$ at time $t$ |
| $T_{set,max}, T_{set,min}$ | maximum and minimum set-point temperatures |
| $T_x^t$ | outdoor (condenser-side) air temperature at time $t$ |
| $k_P, k_I$ | proportional and integral gains of a thermostat controller |
| $t_s, t_e$ | start- and end-hours of the working time in a building |
| $\Delta t$ | unit time step |
| $\Delta t_U$ | time period to update neural networks online |
| $\lambda_y, \lambda_k, \lambda_h$ | weighting factors for HVAC energy cost and constraints for system controllable inputs and system states |
| $\delta_{n,max}$ | maximum value of the $n_{th}$ linear power block |

**C. Variables:**

| | |
|---|---|
| $C_E$ | daily energy cost of HVAC system |
| $J$ | objective value of the reformulated optimization problem with penalty on the system inputs and states |
| $P^t$ | power input to HVAC system at time $t$ |
| $P_c^t$ | power input to HVAC system at time $t$ in a conventional strategy |
| $P_{ref}^t$ | reference power input to HVAC system at time $t$ |
| $Q^t$ | cooling rate supplied by HVAC system at time $t$ |
| $T_{set}^t, T_i^t$ | set-point and indoor temperatures at time $t$ |
| $a_n^t$ | binary variables for piecewise linearization of the variation in $T_i^t$ resulting from HVAC power segment $n$ at time $t$ |
| $e^t$ | difference between $T_{set}^t$ and $T_i^t$ at time $t$ |
| $r_{CR}$ | ratio of reduction in daily energy cost of HVAC system |
| $v_{TC}$ | sum of deviations of $T_i^t$ from an acceptable range during a day |
| $\delta_n^t$ | input power assigned in the linear power block $n$ at time $t$ |

## I. Introduction

COMMERCIAL buildings accounted for more than 36% of the total energy consumption in the United States in 2019 [1]. Heating, ventilation, and air conditioning (HVAC) units represent approximately 40% of the electricity usage in commercial buildings [2]. Therefore, significant attention has been given to the modeling and optimal operation of HVAC systems to improve energy efficiency and electricity bills in commercial buildings.

Physics-based modeling of HVAC units requires numerous parameters to reflect the complex nonlinear relationship among HVAC load, indoor temperatures, and outdoor environments [3]. Therefore, previous studies using simple RC circuit models need further analysis to accurately reflect the thermal dynamics of buildings [4]. Moreover, the types, sizes, and operating characteristics of HVAC systems vary by manufacturer and by the building in which they are installed [5]. This prevents wide applications of physics-based modeling and optimal operation to various buildings with different HVAC systems.

To overcome these challenges, machine learning (ML) is increasingly considered in recent studies on building energy management systems (BEMSs). For example, in [6]–[8], an artificial neural network (ANN) was trained offline via supervised learning (SL) to model the building thermal dynamics. Given the ANN-based model, the solution to the problem for the optimal HVAC system operation was searched for using heuristic algorithms, such as GA, PSO, and firefly algorithms. However, in the heuristic search, the solution is highly likely to fall into one of numerous local optimal solutions. Therefore, the optimization problem should be iteratively solved to find the best solution closer to the global optimum, increasing the computation time [8]. Furthermore, in [6]–[8], only a single ANN was implemented to reflect the highly nonlinear characteristics of the building thermal dynamics. In practice, this has the risk of compromising the modeling accuracy and hence the scheduling performance even for the case of the ANN with deep hidden layers. In [9]–[11], reinforcement learning (RL) was adopted to take the advantage that it requires little knowledge in HVAC system operations and building thermal dynamics. However, in the RL-based search, the optimal solution still needs to be iteratively searched for using an exploration-and-exploitation mechanism, as in the heuristic search. Moreover, the RL can lead to unstable and poor control of an HVAC system particularly in the initial learning episodes [10], implying the difficulty in the RL-based modeling and operation in practice.

The application of ML requires historical data on HVAC system operations under various conditions of building thermal environments. When the size of the data is small and the variability is limited, the ANNs are likely to be over-fitted [12]: i.e., too closely fitted to only a limited set of data points. The requirement of historical data needs to be mitigated for wide applications of the ML-based modeling and optimal operation. For example, in new buildings, insufficient historical data may have been collected yet. Moreover, in traditional energy-inefficient buildings, a rule-based strategy is often adopted to operate HVAC systems with constant or pre-determined temperature set-points. To lessen the data requirement, recent studies have been conducted on online supervised learning. For example, in [13], the optimal operation of an air-conditioning system was achieved online, although variations in the ambient temperature and electricity price were not considered in the operation. In [14], hyper parameters for the optimal HVAC system operation were updated online; however, the temperature set-point was chosen from only a limited set of discrete values and fixed during a day.

This paper proposes a new SL-based strategy for an HVAC system in a commercial building, wherein the optimal temperature set-points are deterministically scheduled using the online SL of interconnected ANNs. Specifically, the system for the building temperature control is divided into three sub-systems: i.e., a thermostat controller, an HVAC unit, and a building envelope. A long short-term memory (LSTM) network is implemented and trained to model each sub-system. The LSTM networks are then interconnected to establish a complete model of the temperature control system. Using the LSTM-based model, an optimization problem is formulated to schedule the optimal temperature set-points, given the day-ahead forecasts of electricity price and thermal environments. The problem is then reformulated, so that the optimal solution can be deterministically searched for using a gradient descent (GD) algorithm. After the optimal scheduling is initiated, the LSTM-based building model continues to undergo online SL, as new data of the building operation are collected. This gradually improves the accuracy of the LSTM models and hence the performance of the optimal scheduling. Case studies are carried out to evaluate the performance of the proposed strategy via comparison with the strategies using a pre-determined temperature set-point, an ideal physic-based building model, and conventional ANN-based building models. The results of the comparative case studies confirm that the proposed strategy ensures the cost-effective operation of the HVAC system and the thermal comfort of occupants. The results also show that the proposed strategy improves the applicability of ML to the modeling and optimal operation of HVAC systems in practice.

The main contributions of this paper are summarized as follows:
• To our best knowledge, this study is the first to develop and interconnect the ANN models of the sub-systems that are required for the building temperature control, mitigating the complexity of the ANNs and improving the modeling accuracy of the building thermal dynamics.
• The interconnected ANNs are directly integrated into the optimization problem for the temperature set-point scheduling. The problem is then reformulated so as to apply a deterministic search algorithm and find the optimal schedule within a reasonable computation time.
• The online SL is incorporated into the optimal scheduling of the HVAC system operation, lessening the requirement of the initial training data and hence facilitating the application of the ML-based modeling and control in practice.
• In the case studies, the proposed strategy is compared with the strategies using a traditional temperature setting rule, an ideal physics-based model, and conventional ANN-based models, enabling a reliable and comprehensive evaluation of the performance of the proposed strategy.

II. MODELING OF BUILDING THERMAL DYNAMICS

A. ANN-based Modeling of Sub-systems

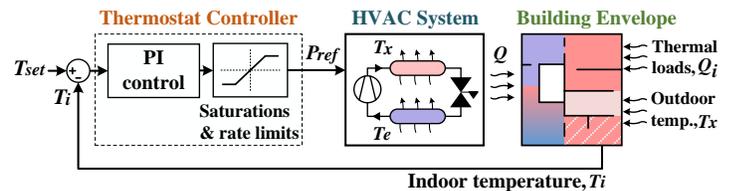

Fig. 1. A schematic diagram of a common system for the building temperature control, consisting of a thermostat control loop, an HVAC unit, and a building envelope.

Fig. 1 shows a common system for the indoor temperate control in a commercial building, which consists of three sub-systems: i.e., a thermostat control loop, an HVAC unit, and a building envelope. Specifically, in the thermostat loop, a proportional-integral (PI) controller is adopted to adjust the reference power input $P^t_{ref}$ of the HVAC unit, based on the difference between the set-point and actual values of the indoor temperature: i.e., $T_{set}^t$ and $T_i^t$, respectively. In practice, the PI controller is accompanied with nonlinear signal processing functions, such as saturations and ramp rate limits, to ensure the reliable HVAC operation. The HVAC unit receives $P^t_{ref}$ as an input signal and provides cooling or heating energy $Q^t$ to the building envelope, given the ambient temperature $T_x^t$ and the evaporator-side air or water temperature $T_e^t$. In this paper, a variable speed heat pump is considered as an example of the HVAC unit [6]. The time response of the variable speed drive is fast and, consequently, the actual power input $P^t$ is almost the same as $P_{ref}^t$, particularly in the scheduling time horizon. In the building envelope, the profile of $T_i^t$ is determined by the HVAC system operation (i.e., $Q^t$) and the building thermal environments $\mathbf{E}^t$, such as $T_x^t$, $T_e^t$, and indoor thermal load $Q_i^t$.

Each sub-system is modeled using an ANN, as shown in Fig. 2. The ANNs are then linked together, based on the interconnections of the sub-systems, discussed above. The operating characteristics of each sub-system can be successfully reflected into an ANN with a rather simple architecture. This mitigates the overall complexity of the ANN model that represents the complete system for the building temperature control, shown in Fig. 1. On the contrary, only a single ANN was often considered in the conventional modeling method [6]–[11]. The ANN then needs to be significantly complicated and deep to accurately reflect the operation of the complete system, requiring a large amount of building operation data. This implies the risk of compromising the modeling accuracy and hence the temperature control performance for the practical case with the limited size and variability of the data.

*B. ANN Architecture and Training*

For the sub-systems, the ANNs are implemented in the form of an LSTM network, which is widely used for time-series data learning. Specifically, the LSTMs consist of multiple hidden layers, each of which includes multiple hidden nodes with self-loops. Furthermore, Fig. 2 shows that the LSTM $\mathbf{L_1}$ has an inner feedback loop between the output and input neurons for $P^t$, which is indicated by the red circles. Similarly, $\mathbf{L_3}$ has an inner feedback loop of $T_i^t$, marked by the yellow circles. An outer feedback loop of $T_i^t$ also exists between the output neuron of $\mathbf{L_3}$ and the input neuron of $\mathbf{L_1}$, which is represented by the blue circles. Moreover, $\mathbf{L_{1-3}}$ have the pre- and post-processors to normalize the input data and recover the output data with their original units, respectively, preventing the training speed from dropping too low.

In addition, each LSTM has a single output neuron and multiple input neurons. As shown in Fig. 2, the outputs of $\mathbf{L_{1-3}}$ are defined as $P^t$, $Q^t$, and $T_i^t$, respectively. The inputs of $\mathbf{L_1}$ are the current and time-delayed values of $T_{set}^t$ and the time-delayed values of $P^t$ and $T_i^t$. For $\mathbf{L_2}$, the inputs are the current and time-delayed values of $P^t$, $T_x^t$, and $T_e^t$. The inputs of $\mathbf{L_3}$ are set to the current and time-delayed values of $Q^t$ and $\mathbf{E}^t$ and the time-delayed $T_i^t$. In this study, the time-delayed inputs of $\mathbf{L_{1-3}}$ are explicitly considered to achieve better accuracy in modeling the building thermal dynamics by reflecting the effects from the integral controller in the thermostat loop, the heat exchanger in the HVAC system, and the thermal energy storage inherent in the building envelope, respectively. Note that the length of each time-delayed dataset is set to $L_P$, which is determined considering the trade-off between the modeling accuracy and computational burden.

The individual LSTMs are trained separately using the database of a BEMS to determine the weighting coefficients and biases for all the input, hidden, and output neurons. During the training, the feedback loops are open to feed the actual time-delayed data into the input neurons. The LSTMs are then interconnected and tested with closed feedback loops, so that the outputs estimated at the current time step can be used as the time-delayed inputs at the next step. The LSTMs are integrated into the optimization problem for the day-ahead set-point scheduling, as discussed in Section III. In other words, the physics-based modeling coefficients of the HVAC system and the building envelope are not required for the optimal scheduling, enabling the wide application of the proposed strategy in practical BEMSs.

III. Optimal Scheduling Integrated with Online SL

*A. Optimization Problem Formulation*

Using the trained $\mathbf{L_{1-3}}$, the optimal schedule of $T_{set}^t$ can be determined by solving $\mathbf{S_1}$ as:

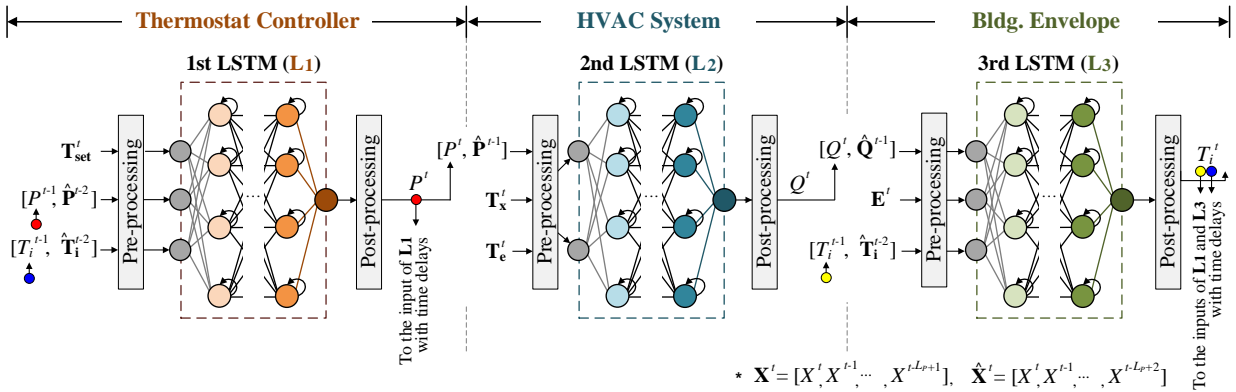

Fig. 2. Interconnection of the LSTM networks that correspond to the models of the thermostat controller, HVAC system, and building envelope, respectively.

**S₁: Problem for optimal HVAC system operation**

$$\arg\min_{T_{set}^t} \quad C_E = \sum_{t=1}^{N_T} C^t P^t, \quad (1)$$

subject to
$$T_{set,\min} \leq T_{set}^t \leq T_{set,\max}, \quad \forall t, \quad (2)$$
$$T_{i,\min}^t \leq T_i^t \leq T_{i,\max}^t, \quad \forall t, \quad (3)$$
$$P_{\min} \leq P^t \leq P_{\max}, \quad \forall t, \quad (4)$$
$$R_L \leq (P^t - P^{t-\Delta t})/\Delta t \leq R_H, \quad \forall t, \quad (5)$$

where
$$P^t = \mathbf{L_1}\left(T_{set}^t, \cdots, T_{set}^{t-L_p+1}, P^{t-1}, \cdots, P^{t-L_p}, T_i^{t-1}, \cdots, T_i^{t-L_p}\right), \quad \forall t, \quad (6)$$
$$Q^t = \mathbf{L_2}\left(P^t, \cdots, P^{t-L_p+1}, T_x^t, \cdots, T_x^{t-L_p+1}, T_e^t, \cdots, T_e^{t-L_p+1}\right), \quad \forall t, \quad (7)$$
$$T_i^t = \mathbf{L_3}\left(Q^t, \cdots, Q^{t-L_p+1}, \mathbf{E}^t, \cdots, \mathbf{E}^{t-L_p+1}, T_i^{t-1}, \cdots, T_i^{t-L_p}\right), \quad \forall t. \quad (8)$$

The objective function (1) aims to minimize the energy cost $C_E$ of the HVAC system: i.e., the 24-h sum of the hourly-varying retail electricity price $C^t$ multiplied by the power input $P^t$ of the HVAC system. Note that $C^t$ can be negative, for example, when there is an excess of renewable generation [15]. The optimal schedule of $T_{set}^t$ is determined using day-ahead forecasts of the electricity price and weather conditions. In this study, the forecast data are assumed to be already available in the BEMS database for brevity, as in [7], [8] and [16]; the integration of price and weather forecasting algorithms has been left for future research.

In the set of constraints, (2) shows the limits of the operating range of the thermostat controller (i.e., from $T_{set,min}$ = 15°C to $T_{set,max}$ = 35°C) to secure the reliable operation of the HVAC system. Moreover, (3) represents that $T_i^t$ should be maintained within an acceptable range from $T_{i,min}^t$ to $T_{i,max}^t$ to ensure the thermal comfort of occupants. Note that $T_{set}^t$ and $T_i^t$ can differ under normal operating conditions of the HVAC system mainly due to the large thermal capacity in the building envelope. The constraints (4) require $P^t$ to be maintained between $P_{max}$ and $P_{min}$; in this paper, these are set to the rated power input and zero, respectively. Furthermore, (5) specifies the limits on the upward and downward ramp rates of $P^t$ for the time period of $\Delta t$ = 1 h. In (5), $P^t$ at $t = 0$ h is set to zero, assuming that the HVAC system is turned off at night (after 7 pm to midnight) when the commercial building has low occupancy. In (6)–(8), the LSTM-based sub-system models, discussed in Section II, are parameterized as the functions $\mathbf{L_{1-3}}(\cdot)$, in which the current and time-delayed inputs and the output are specified.

The optimization problem $\mathbf{S_1}$ [i.e., (1)–(8)] can be expressed in a compact form as:

**S₂: Compact form of the original problem S₁**

$$\arg\min_{u^t} \quad C_E = \sum_{t=1}^{N_T} y^t, \quad (9)$$

subject to
$$u_{\min} \leq u^t \leq u_{\max}, \quad \forall t, \quad (10)$$
$$\mathbf{s}_{\min}^t \leq \mathbf{s}^t \leq \mathbf{s}_{\max}^t, \quad \forall t, \quad (11)$$
$$y^t = f(\mathbf{s}^t, \mathbf{v}^t), \quad \forall t, \quad (12)$$
$$\mathbf{s}^t = g(\mathbf{s}^{t-1}, \mathbf{v}^t), \quad \forall t, \quad (13)$$

where $y^t = C^t \cdot P^t$ and $u^t = T_{set}^t$ are defined as the output and the controllable input, respectively, of the system for the optimal building temperature control. For notational simplicity, a variable vector $\mathbf{v}^t$ is used to represent the system inputs $[u^t, \mathbf{w}^t]^T$, including the system disturbances $\mathbf{w}^t = \mathbf{E}^t$. Similarly, $\mathbf{s}^t$ indicates the system states at time $t$: i.e., $\mathbf{s}^t = [T_i^t, Q^t, P^t, \Delta P^t]^T$.

To deterministically search for the optimal $u^t$ during $1 \leq t \leq N_T$, $\mathbf{S_2}$ is then reformulated to $\mathbf{S_3}$ as:

**S₃: Reformulated problem of S₂**

$$\arg\min_{u^t} J = \sum_{t=1}^{N_T}\left\{\lambda_y (y^t)^2 + \lambda_k (k^t)^2 + \lambda_h (\mathbf{h}^t)^T \cdot \mathbf{h}^t\right\}, \quad (14)$$

where
$$k^t = \begin{cases} u^t - u_{\max} & \text{for } u^t > u_{\max} \\ u_{\min} - u^t & \text{for } u^t < u_{\min}, \quad \forall t, \\ 0 & \text{otherwise} \end{cases} \quad (15)$$

$$\mathbf{h}^t = \begin{cases} \mathbf{s}^t - \mathbf{s}_{\max}^t & \text{for } \mathbf{s}^t > \mathbf{s}_{\max}^t \\ \mathbf{s}_{\min}^t - \mathbf{s}^t & \text{for } \mathbf{s}^t < \mathbf{s}_{\min}^t, \quad \forall t, \\ 0 & \text{otherwise} \end{cases} \quad (16)$$

(12) and (13).

Specifically, in (14), the objective function is implemented using a quadratic function of the HVAC energy cost (i.e., $(y^t)^2 = (C^t \cdot P^t)^2$). Moreover, (10) and (11) are relaxed to (15) and (16), respectively, and then added to (14) in a quadratic form of the penalties that are incurred when the constraints on $u^t$ and $\mathbf{s}^t$ are violated; in (14), $\lambda_k$ and $\lambda_h$ are the corresponding penalty factors. As shown in (12)–(16), $\mathbf{S_3}$ is a continuous, nonlinear function and hence its minimum can readily be found using a deterministic search algorithm. In this paper, a GD algorithm [17], [18] is adopted for the optimal search, where the next step is determined proportional to the negative gradient of $\mathbf{S_3}$ at the current step. Since it requires only the first derivative, the GD solver can readily be implemented in the BEMS, facilitating the optimal operation of the HVAC system in practice.

*B. Online Supervised Learning*

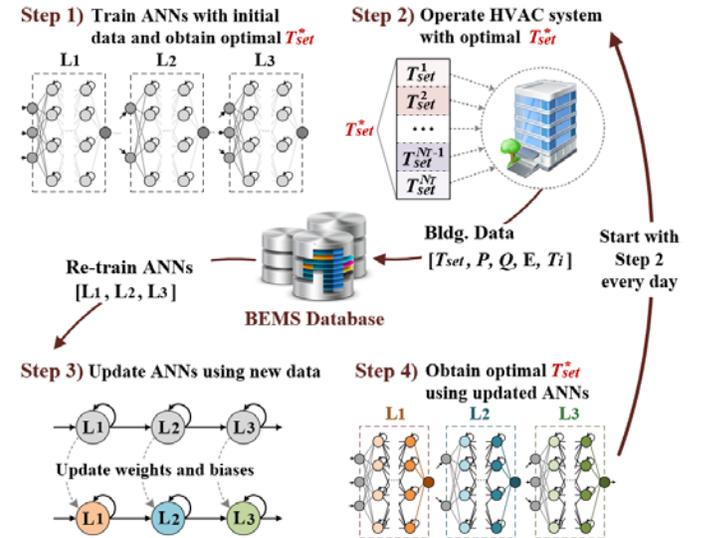

Fig. 3. Flowchart for the online SL of $\mathbf{L_{1-3}}$ that is integrated with the optimal scheduling of the HVAC system.

After the optimal scheduling of $T_{set}^t$ is initiated, $\mathbf{L_{1-3}}$ undergo repeated online SL, as new data of $T_{set}^t$, $P^t$, $Q^t$, and $T_i^t$ are obtained under various profiles of $C^t$ and $\mathbf{E}^t$. This gradually mitigates the overfitting of $\mathbf{L_{1-3}}$. In other words, $\mathbf{L_{1-3}}$ become well adapted to changes in the operating conditions of the building, improving the accuracy in modeling the building thermal dynamics. Specifically, Fig. 3 shows a flowchart for the online SL of $\mathbf{L_{1-3}}$. In Step 1, the optimal day-ahead scheduling of $T_{set}^t$ is initiated, after $\mathbf{L_{1-3}}$ are trained with the initial historical data of the BEMS. Due to the small size and variability of the data, $\mathbf{L_{1-3}}$ are likely to be rather inaccurate, limiting the performance of the optimal scheduling. In Step 2, the HVAC system operates according to the optimal schedule of $T_{set}^t$ on day $d$, and the BEMS collects the corresponding dataset [$T_{set}^t$, $P^t$, $Q^t$, $\mathbf{E}^t$, $T_i^t$] for $1 \leq t \leq N_T$. The profiles of the dataset are likely to be different from those of the historical BEMS datasets before the optimal scheduling is initiated. This increases the variability of the training data, improving the accuracy of $\mathbf{L_{1-3}}$ when they are re-trained using the newly collected dataset in Step 3. In Step 4, the optimization problem $\mathbf{S_3}$ is updated using the retrained $\mathbf{L_{1-3}}$ and solved for the forecasts of $\mathbf{E}^t$ on day $d$+1. The improved accuracy of $\mathbf{L_{1-3}}$ will lead to the expansion of the feasible solution area of $\mathbf{S_3}$, enhancing the performance of the optimal HVAC system operation. Step 2 is then repeated on day $d$+1 with the new optimal schedule of $T_{set}^t$. In this paper, Steps 2–4 are performed in every period of the scheduling day to achieve a continuous improvement of the modeling accuracy and the scheduling performance.

## IV. CASE STUDIES AND SIMULATION RESULTS

### A. Test Conditions

The proposed strategy was tested for an experimental setup of an office building with an HVAC system, shown in Fig. 4. Briefly, the experimental setup is divided into test and climate rooms, both of which are within a larger laboratory room with the temperature of $T_a^t$. The test room has lights and heat sources to emulate the internal thermal load $Q_i^t$ in a common office. The walls and floor consist of multiple layers of different building materials. For the case studies, the power rating of the HVAC system was set to $P_{max}$ = 50 kW, and a scaled faction of the corresponding $Q^t$ was used to control $T_i^t$ in the test room. The climate room contains a separate heating unit to emulate the building thermal environments $\mathbf{E}^t$. For the experimental setup, a building simulator was implemented in [19] to estimate $T_i^t$ for $P^t$, given $\mathbf{E}^t = [T_x^t, T_a^t, T_e^t, Q_i^t]$. In this study, the simulator was further extended by integrating the thermostat control loop with the HVAC system, as shown in Fig. 1; this enabled the indirect control of the HVAC unit, as in common practical buildings.

To establish the initial training datasets, the building simulator was run using the data of $Q_i^t$ estimated from a real building [19], [20] and of $T_x^t$ measured in Boston from June 1 to August 31 of 2017–2019 [21], as shown in Fig. 5(a) and (b), respectively. Note that $Q_i^t$ also can be surveyed and measured for benchmark buildings [22]. Given $Q_i^t$ and $T_x^t$, the simulator was run with the pre-determined profiles of $T_{set}^t$, such that $T_i^t$ was controlled within an acceptable range under the condition of the traditional HVAC system operation. Fig. 5(c) shows the corresponding profiles of $P^t$ obtained from the simulation runs. Moreover, Fig. 5(d) shows the profiles of $C^t$ [23] for the same time period where the data of $Q_i^t$ and $T_x^t$ were acquired. Note that on several days, $C^t$ decreased below zero in the early morning. The size of the initial datasets [$T_{set}^t$, $P^t$, $Q^t$, $\mathbf{E}^t$, $T_i^t$] was 1,200 (i.e., 50 days) and 8 with respect to time and objects, respectively. The size with respect to time continued to increase, as the optimal profiles of $T_{set}^t$, $P^t$, $Q^t$, and $T_i^t$ were obtained from $\mathbf{S_3}$ for days $d$ = 1 to $N_d$ (i.e., 200), as discussed in Section III-B. In other words, the online SL was conducted to train $\mathbf{L_{1-3}}$ and solve $\mathbf{S_3}$ during the period from $d$ = 1 to $N_d$. Note that the time-delayed data for the objects were not considered in the size estimation. The datasets were then randomly shuffled and divided into three parts with the ratios of 0.8:0.1:0.1 for the training, validation, and testing, respectively. Moreover, Table I lists the parameter values used for the proposed strategy. In particular, the learning rates $R_{O1-3}$ were reduced from $10^{-3}$ to $10^{-4}$ when the epoch number increased greater than two thirds of the total number of epochs. This aimed to achieve high accuracy of $\mathbf{L_{1-3}}$ and hence of the optimal solutions to $\mathbf{S_3}$.

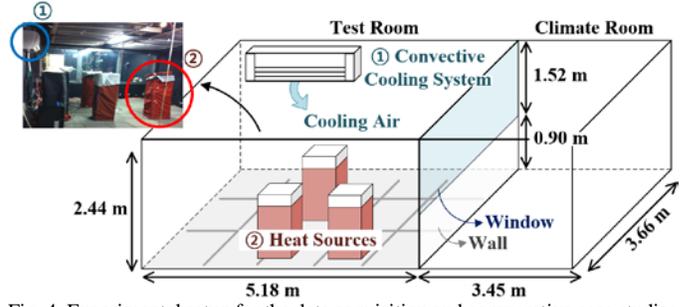

Fig. 4. Experimental setup for the data acquisition and comparative case studies

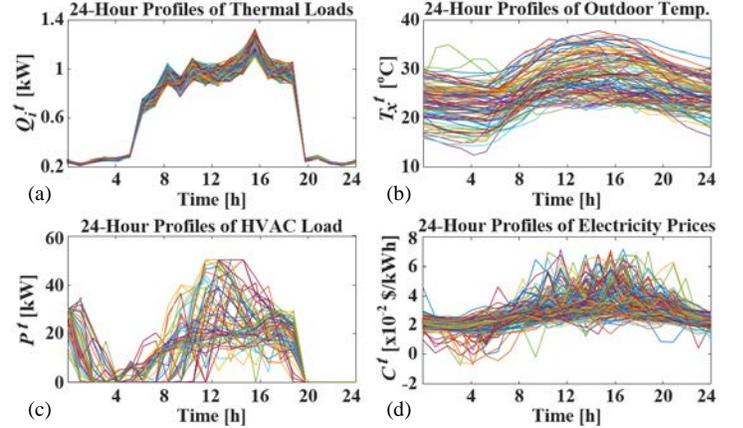

Fig. 5. Case study conditions from June 1 to August 31, 2017: (a) $Q_i^t$, (b) $T_x^t$, (c) $P^t$, and (d) $C_t$. The profiles of $Q_i^t$, $T_x^t$, $P^t$, and $C^t$ in 2018 and 2019 were similar.

TABLE I. PARAMETER VALUES FOR THE CASE STUDIES

| Parameters | Values | Units | Parameters | Values |
|---|---|---|---|---|
| $P_{min}$, $P_{max}$ | 0, 50 | [kW] | $N_d$ | 200 |
| $R_L$, $R_H$ | –40, 30 | | $N_S$ | 4 |
| $L_P$ | 24 | [h] | $N_{ID}$ | 1200 |
| $N_T$ | 24 | | $N_{HL1}$, $N_{HL2}$, $N_{HL3}$ | 3, 4, 3 |
| $\Delta t$, $\Delta t_U$ | 1, 24 | | $N_{HN1}$, $N_{HN2}$, $N_{HN3}$ | 15, 20, 20 |
| $t_s$, $t_e$ | 7, 19 | | $N_{ET}$, $N_{EO}$ | 5000, 1000 |
| $T_{i,min}^t$, $T_{i,max}^t$ | 22, 24 | [°C] | $R_T$ | $4 \times 10^{-3}$ |
| $T_{set,min}$, $T_{set,max}$ | 15, 35 | | $R_{O1}$, $R_{O2}$, $R_{O2}$ | $10^{-3}$, $10^{-4}$ |
| $D_T$ | 1 | | $\lambda_y$, $\lambda_k$, $\lambda_h$ | 2.0, 0.5, 7.5 |

TABLE II. FEATURES OF THE PROPOSED, IDEAL, AND RULE-BASED STRATEGIES

| Strategies | | Set-point temperatures | Building modeling | Opt. solver | Online learning |
|---|---|---|---|---|---|
| Proposed | Case 1 | optimized | actual data of building operation | GD | ○ |
| Ideal | Case 2 | optimized | fully-informed model parameters | MILP | - |
| Rule-based | Case 3 | pre-determined | - | - | - |

The HVAC system operations were compared for three cases: i.e., the proposed SL-based strategy (Case 1), an ideal physics-based strategy (Case 2), and a traditional rule-based strategy (Case 3). Table II lists the main features of Cases 1–3. In Case 2, the piecewise linear equations for variations in $T_i^t$ for a change in $P^t$ were established using the complete information on the physics-based modeling parameters of the HVAC system and building envelope [24]; see (A1)–(A5) in the Appendix. The optimal schedule of $T_{set}^t$ was then obtained by replacing (6)–(8) with (A1)–(A5) and then applying mixed-integer linear programming (MILP). Note that Case 2 is referred to as the ideal case, because most of the information is not available in practice. In Case 3, $T_{set}^t$ was fixed at 23°C, regardless of the variation in $C_t$. For the fair comparison of Cases 1–3, the HVAC system was assumed to be capable of operating from $t = 1$ h in the case studies. This also allowed the building to take the advantage of pre-cooling for all Cases 1–3; the late start of the HVAC operation has the risk of causing an increase in $C_E$ and a deviation of $T_i^t$ from the acceptable range.

### B. Improvement via Online Supervised Learning

The accuracy of the LSTM-based building model was verified by comparing the actual values of $P^t$, $Q^t$, and $T_i^t$ in the testing datasets (discussed in Section IV-A) with the corresponding estimates obtained from $\mathbf{L_{1-3}}$. Note that the estimates were acquired after training and interconnecting $\mathbf{L_{1-3}}$. Fig. 6 shows the results of the comparisons for $d = 1$ and $N_d$, where the x- and y-axes represent the actual values and the estimates, respectively. For $d = 1$, the normalized root mean square errors (nRMSEs) of $\mathbf{L_{1-3}}$ were estimated to be rather considerable: i.e., $1.1 \times 10^{-1}$, $9.5 \times 10^{-3}$, and $5.8 \times 10^{-3}$, respectively. However, as the online SL and optimal scheduling continued, the nRMSEs for $d = N_d$ were reduced to $9.1 \times 10^{-3}$, $3.1 \times 10^{-3}$, and $4.8 \times 10^{-4}$, respectively. Fig. 7 shows the variations in the nRMSEs over the time period from $d = 1$ to $N_d$. For all $\mathbf{L_{1-3}}$, the nRMSEs were reduced rapidly during the initial period and decreased gradually for the remaining period. The results of the case studies confirmed that the online SL integrated with the optimal scheduling is effective in improving the accuracy of the LSTM-based models of the sub-systems (and hence the complete system) for the building temperature control.

In addition, Table III and Fig. 8 show the optimal scheduling results for the proposed strategy (i.e., Case 1) in comparison with those for the ideal and traditional strategies (i.e., Cases 2 and 3). Specifically, Table III lists the average value of the reduction rates of $C_E$ for Case 1, compared to $C_E$ for Case 3, during the time period of every $N_d/4$ (i.e., 50) days. It can be seen that as the online SL continued, the average reduction rate gradually increased from 20.96% to 23.36%. Fig. 8(a) shows the comparisons of $C_E$ for Cases 1–3 for each day $d$. For Case 1, $C_E$ was only slightly larger than for Case 2 but considerably smaller than for Case 3. Moreover, Fig. 8(b) shows the average of the accumulated deviations of $T_i^t$ during the period of every $N_d/4$ days, given by:

$$v_{TC} := \sum_{t=1}^{N_T} \left\{ \max\left(T_i^t - T_{i,\max}^t, 0\right) + \max\left(T_{i,\min}^t - T_i^t, 0\right) \right\}. \quad (17)$$

It can be seen that the average of $v_{TC}$ for Case 1 was gradually reduced and became comparable with that for Case 2. As $\lambda_h$ in (14) increases, $v_{TC}$ can be reduced more rapidly and maintained further lower, although $C_E$ is likely to increase. The results of the case studies confirm that the proposed strategy is effective in reducing the HVAC energy cost, while ensuring occupants' thermal comfort. The proposed strategy is more practically applicable than existing RL-based strategies, in which the HVAC system operation is likely to be unstable and the occupants' thermal comfort is expected to be unsatisfied in the initial learning episodes.

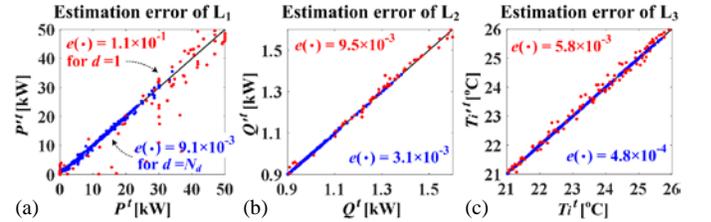

Fig. 6. Comparisons of the actual and estimated values of $P^t$, $Q^t$, and $T_i^t$. The red and blue dots indicate the test results of $\mathbf{L_{1-3}}$ for $d = 1$ and $N_d$, respectively.

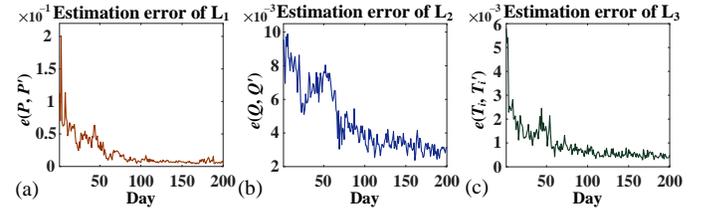

Fig. 7. Variations in the nRMSEs of $\mathbf{L_{1-3}}$ for the time period from $d = 1$ to $N_d$.

TABLE III. AVERAGE VALUES OF THE COST REDUCTION RATES FOR CASE 1

| Periods | $0 < d \leq N_d/4$ | $N_d/4 < d \leq N_d/2$ | $N_d/2 < d \leq 3N_d/4$ | $3N_d/4 < d \leq N_d$ |
|---|---|---|---|---|
| $r_{CR}$ [%] | 20.96 | 21.66 | 22.62 | 23.36 |

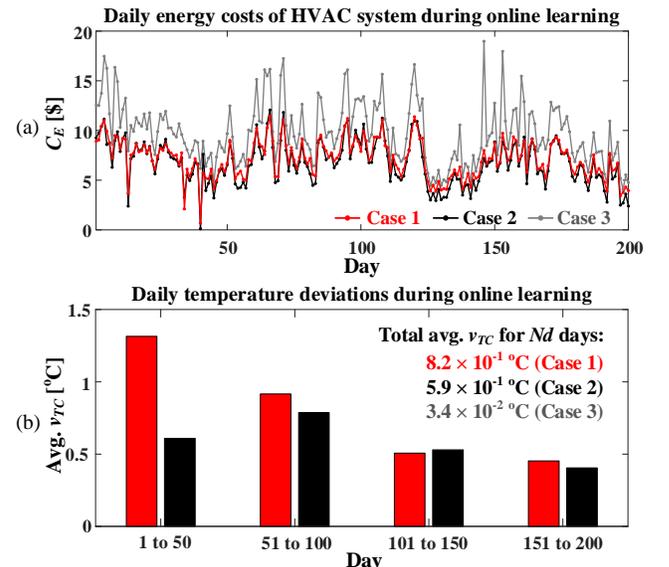

Fig. 8. Optimal scheduling results for Cases 1–3 over the time period from $d = 1$ to $N_d$: (a) $C_E$ and (b) the average of $v_{TC}$.

## C. Comparisons of Operating Schedules of HVAC System

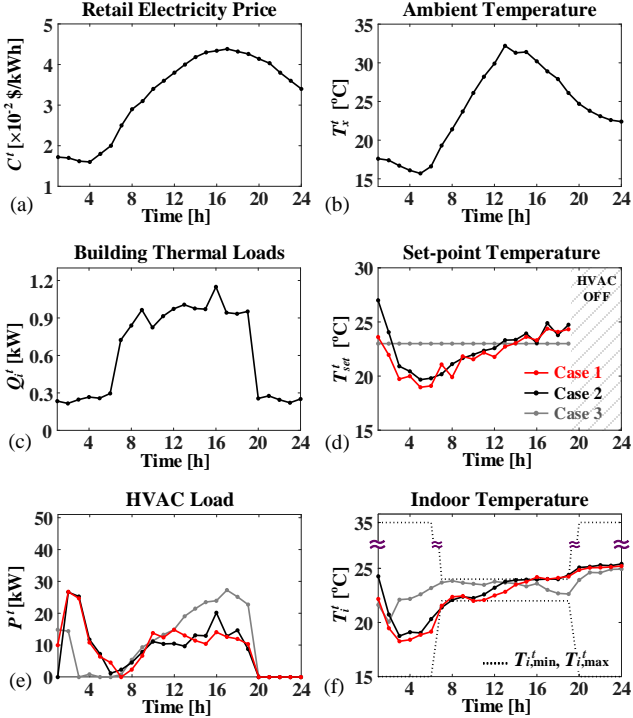

Fig. 9. Comparisons of the 24-h schedules for Cases 1–3: (a) $C^t$, (b) $T_x^t$, (c) $Q_i^t$, (d) $T_{set}^t$, (e) $P^t$, and (f) $T_i^t$.

TABLE IV. COMPARISONS OF THE HVAC ENERGY COST,
COST REDUCTION RATES, AND OCCUPANTS' THERMAL DISCOMFORT

| Profiles of $C^t$ and $\mathbf{E}^t$ | | Proposed (Case 1) | Ideal (Case 2) | Rule-based (Case 3) |
|---|---|---|---|---|
| Fig. 9 | $C_E$ [$] | 6.74 | 6.74 | 9.39 |
| | $r_{CR}$ [%] | 28.2 | 28.2 | - |
| | $v_{TC}$ [ºC] | 1.03 | 1.03 | 0 |
| Fig. 10 | $C_E$ [$] | 4.22 | 4.25 | 7.42 |
| | $r_{CR}$ [%] | 43.1 | 42.7 | - |
| | $v_{TC}$ [ºC] | 0.63 | 0.52 | 0 |

Fig. 9 represents the 24-h schedules of $T_{set}^t$ and the corresponding variations in $P^t$ and $T_i^t$ for Cases 1, 2, and 3, given the forecasts of $C^t$ and $\mathbf{E}^t$. Specifically, for Case 1, $T_{set}^t$ was scheduled at relatively low levels in the early morning due to the low values of $C^t$, whereas $T_x^t$ and $Q_i^t$ were maintained high during $7\,h \leq t \leq 19\,h$. As $C^t$ began to increase, $P^t$ for Case 1 then became lower than that for Case 3. In other words, the proposed strategy achieved the HVAC load shift from on-peak hours to off-peak hours, leading to the pre-cooling operation and hence the reduction of the HVAC energy cost. Table IV shows that $C_E$ for Case 1 was estimated as $6.74, which is 28.2% less than $9.39 for Case 3. Fig. 9(f) shows that in Case 1, $T_i^t$ was still successfully controlled within the acceptable range.

Fig. 10 shows the scheduling results for different profiles of $C^t$ and $\mathbf{E}^t$. Specifically, $C^t$ differed more between the off- and on-peak hours. Fig. 10(a) shows that $C^t$ was negative at $t = 3\,h$ and 5 h and increased up to 12.1 ₵/kWh at $t = 15\,h$; note that the y-axis was broken to better display the variation in $C^t$. Moreover, $Q_i^t$ was estimated lower, compared to the case for Fig. 9(c). Therefore, the shift of $P^t$ became larger than the case for Fig. 9(e), leading to a larger reduction in $C_E$: i.e., from $r_{CR} =$ 28.2% to 43.1%. This confirms that the proposed strategy succ-essfully reflects the load shifting capabilities of the HVAC system in response to the time-varying electricity prices and building thermal conditions.

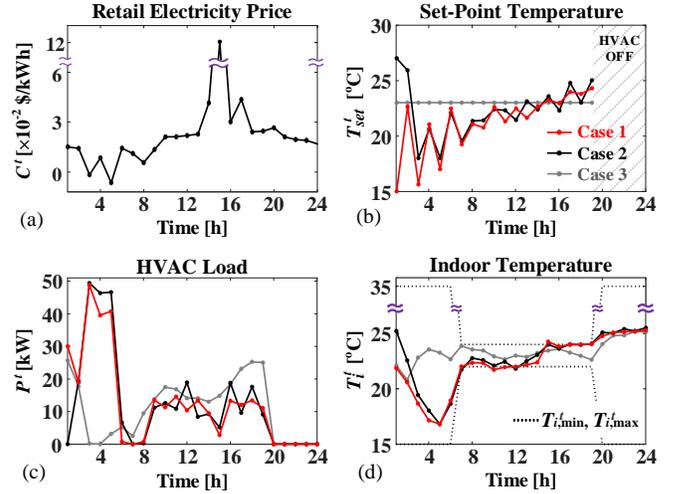

Fig. 10. Comparisons of the results for the proposed, ideal, and conventional strategies for different profiles of $C^t$ and $\mathbf{E}^t$: (a) $C^t$, (b) $T_{set}^t$, (c) $P^t$, and (d) $T_i^t$. In Fig. 10(a), the y-axis is broken to accommodate the peak of $C^t$ for $t = 15\,h$.

In Figs. 9 and 10, the optimal schedules for the proposed and ideal strategies (i.e., Cases 1 and 2) were considerably similar, confirming the accuracy of $\mathbf{L}_{1-3}$. The small difference was mainly because the proposed strategy was developed using the actual operating data of the temperature control system, whereas the ideal strategy was achieved using the complete information on the system modeling parameters. It was also attributable to the difference between the GD and MILP solvers used for the proposed and ideal strategies, respectively.

## D. Comparisons with Conventional SL-based Strategies

TABLE V. COMPARISONS BETWEEN
THE PROPOSED AND CONVENTIONAL SL-BASED STRATEGIES

| Profiles of $C^t$ and $\mathbf{E}^t$ | | Online SL | | | Offline SL |
|---|---|---|---|---|---|
| | | 3 LSTMs | 2 LSTMs | 1 LSTM | 1 LSTM |
| Fig. 9 | $e(T_i, T_i')$ | $4.02 \times 10^{-4}$ | $1.53 \times 10^{-3}$ | $3.71 \times 10^{-3}$ | $3.86 \times 10^{-2}$ |
| | $C_E$ [$] | 6.74 | 7.11 | 8.11 | 10.6 |
| | $r_{CR}$ [%] | 28.2 | 31.6 | 22.0 | -1.92 |
| | $v_{TC}$ [ºC] | 1.03 | 1.20 | 1.34 | 0.55 |
| | comp. time [s] | 1,618 | 1,377 | 1,052 | 1,073 |
| Fig. 10 | $e(T_i, T_i')$ | $3.97 \times 10^{-4}$ | $1.24 \times 10^{-3}$ | $7.09 \times 10^{-3}$ | $3.82 \times 10^{-2}$ |
| | $C_E$ [$] | 4.22 | 5.00 | 5.20 | 7.66 |
| | $r_{CR}$ [%] | 43.1 | 32.6 | 29.9 | 3.23 |
| | $v_{TC}$ [ºC] | 0.63 | 0.94 | 1.23 | 0 |
| | comp. time [s] | 1,616 | 1,441 | 1,054 | 1,074 |

The case studies discussed in Sections IV-C and IV-D were repeated to further evaluate the performance of the proposed SL-based strategy. In particular, as shown in Table V, the evaluation was achieved via the comparison with the conventional SL-based strategies, in which a single LSTM was trained offline and online to model the temperature control system. The case with the two online-trained LSTMs was also considered, the first of which modeled the thermostat control loop, and the second corresponded to the HVAC system and the building envelope. The results of the comparative case studies confirm that the proposed strategy is more effective in improving the building modeling accuracy and the temperature control performance, while maintaining the computation time within reasonable limits. The computation time was estimated on a computer with a six-core 4.3-GHz CPU and 32 GB of RAM.

## V. Conclusions

This paper proposed a new SL-based strategy for the optimal operation of an HVAC system in a commercial building. The system for the indoor temperature control was divided into three sub-systems, each of which is modeled using an LSTM. The LSTMs were then interconnected and integrated directly into the optimization problem for the temperature set-point scheduling. The optimization problem was reformulated and solved using a deterministic search algorithm within a reasonable computation time. After the optimal scheduling was initiated, the interconnected LSTMs went through the online SL repeatedly, gradually improving the modeling accuracy and the scheduling performance. The case studies were conducted to validate the performance of the proposed strategy in comparison with the strategies using the rule-based temperature set-point, the ideal physics-based building model, and the conventional ANN-based building models. The case study results confirmed that the proposed strategy accurately reflects the load-shifting capability of the HVAC system in response to the time-varying conditions of the electricity price and building thermal environments, successfully reducing the HVAC energy cost. The results also verified that the proposed strategy effectively mitigates the requirement of the historical building data and the risk of an unstable operation of the HVAC system and the thermal discomfort of occupants in the initial learning period, which is of the utmost importance for the practical application in BEMSs.

## Appendix

For the comparative case studies, a physics-based model of the system for the building temperature control was implemented as:

$$e^t = T_{set}^t - T_i^{t-1}, \quad \forall t, \tag{A1}$$

$$P_{ref}^t = P_O - k_P e^t - k_I \sum_{\tau=1}^{t} e^\tau, \quad \forall t, \tag{A2}$$

$$\sum_{n=1}^{N_S} \delta_n^t = P^t, \quad \forall t, \tag{A3}$$

$$T_i^t = T_{in}^t + \sum_{\tau=1}^{t} \sum_{n=1}^{N_S} F_{n,\tau}^t \delta_n^\tau, \quad \forall t, \tag{A4}$$

$$\delta_{n,\max} a_n^t \leq \delta_n^t \leq \delta_{n,\max} a_{n-1}^t, \quad a_0^t = 1, \quad a_{N_S}^t = 0,$$
$$\forall a_n^t \in \{0,1\}, \forall n, \forall t. \tag{A5}$$

The constraints (A1) and (A2) represent the operation of the PI controller in the thermostat feedback loop. Moreover, (A3)–(A5) correspond to the piecewise linear approximation of the nonlinear variation in $T_i^t$ for a change in $P^t$ [24]. Specifically, in (A3), $P^t$ is divided into $N_S$ linear blocks. In (A4), the variation from $T_{in}^t$ to $T_i^t$ is calculated as the sum of the temperature variations that are led by the incremental HVAC loads assigned in the linear blocks. This is possible, because in (A4), $F_{n,\tau}^t$ contains the complete information on the inter-time thermal response of the building to the HVAC system operation. Moreover, (A5) represents the boundaries of the linear blocks to complete the piecewise linear approximation.